# Light and Sound Driven Wavefront Shaping and Imaging through Scattering Tissue

Fei Xia[1,2,3,†], Ivo Leite[2,†,], Shuquan Xiao[1,2], Nikita Kaydanov[2], Frederik Goerlitz[2], Sylvain Gigan[1,*], Robert Prevedel [2,*]

[1] Laboratoire Kastler Brossel, ENS-Universite PSL, CNRS, Sorbonne Université, Collège de France, 24 Rue Lhomond, F-75005 Paris, France
[2] European Molecular Biology Laboratory, Meyerhofstraße 1, 69117 Heidelberg, Germany
[3] Nhu Department of Electrical Engineering and Computer Science, University of California, Irvine, 92697, Irvine, USA
[†] These authors contributed equally
[*] Corresponding authors: S.G.: sylvain.gigan@lkb.ens.fr, R.P.: prevedel@embl.de

## Abstract

Deep, high-resolution imaging is essential for unraveling biological complexity and advancing medical diagnostics, yet scattering fundamentally limits optical methods. Among the most promising approaches, photoacoustic imaging achieves penetration into deep tissue but with coarse resolution, while fluorescence provides subcellular detail but is confined to shallow depths. This depth-resolution trade-off remains a central barrier to biomedical imaging. To bridge this fundamental gap, we present a hybrid dual-modal strategy that combines the benefits of photoacoustic and fluorescence modalities. Our approach leverages hybrid opto-acoustic feedback for wavefront shaping and computational imaging through scattering media. By combining these complementary signals into a nonlinear feedback metric, we achieve robust optical focusing even under signal degradation. In particular, we show that photoacoustic-guided wavefront shaping inherently generates fluorescence that can be harvested for computational high-resolution imaging even within highly scattering biological tissues, thereby leveraging the complementary strengths of both modalities in a single framework. Proof-of-concept experiments demonstrate this synergistic approach, paving the way for optical imaging techniques that fully leverage the potential of such dual-modalities for large depth penetration and high resolution in complex biological tissues.

## Introduction

Deep optical imaging is fundamental for understanding biological organization and function across multiple spatial scales, from subcellular structures to intact tissues and organisms[1,2]. However, light scattering within biological tissue fundamentally limits both penetration depth and spatial resolution[1,3]. As photons undergo multiple scattering events, optical focus and contrast deteriorate, creating a depth-resolution trade-off that has long restricted the ability of optical methods to visualize complex biological systems in situ[1,4].

Among established techniques, photoacoustic (PA) imaging[5–9] and fluorescence microscopy[10–13] are two of the most widely used and complementary modalities. PA imaging converts absorbed optical energy into ultrasound waves, and thanks to the weak scattering of acoustic waves in tissues, achieves millimeter- to centimeter-scale penetration with strong optical absorption contrast[5–9]. Its spatial resolution, however, is constrained by the acoustic wavelength, typically well beyond tens of micrometers up to millimeters[9]. In contrast, fluorescence microscopy offers molecular specificity and diffraction-limited (<1µm), subcellular resolution but rapidly loses signal beyond superficial depths due to scattering and absorption[1]. Despite major advances in both fields such as multispectral PA tomography[14,15], adaptive optics[3,16–19], and multiphoton excitation[20–22], the intrinsic trade-off between imaging depth and resolution remains unresolved and has so far prevented the ability to capture high-resolution images inside deep biological tissues[1,2,4].

Wavefront shaping has emerged as a promising approach to mitigate scattering by actively controlling the optical phase of the incident light field[3,23,24]. Because light propagation through scattering media is deterministic[24–26], a properly tailored input wavefront can, in principle, interfere constructively at a target location, effectively reconstituting an optical focus[1,3,23]. Feedback-based wavefront shaping iteratively optimizes the input phase using a measured feedback signal, commonly fluorescence signal[27] or photoacoustic signal[28–32], to enhance light delivery[33]. This strategy has demonstrated impressive focusing performance through scattering media[3,26,33–35]. However, practical implementation in biological tissue is hindered by several factors: the feedback signals are often weak[1], linearly dependent on local intensity[36], and highly sensitive to noise[3]. As a result, conventional single-modality feedback methods struggle to maintain stable focusing under strong scattering or deep-tissue conditions[4].

Parallel to these hardware-based efforts, computational fluorescence imaging approaches have been developed more recently. These generally seek to recover fluorescent structures algorithmically from scattered photons[27,37–39]. Instead of physically refocusing light, these approaches reconstruct fluorescence distributions using matrix inversion[27] and speckle correlation[38–40]. More recent strategies incorporate structured illumination[41], robust principle component analysis[39], or machine-learning-based reconstruction[42] to improve image quality hidden behind scattering layers. Despite these developments, computational fluorescence imaging remains fundamentally limited by the rapid degradation of the excitation field in deep layers and the absence of a reliable, high-SNR feedback mechanism[1,27,43]. As scattering increases, excitation light becomes spatially diffuse, reducing both the confinement of fluorescence generation and the stability of the inversion process[1,2]. Consequently, current computational fluorescence methods are largely confined to thin or weakly scattering samples and cannot achieve deep or high-fidelity imaging in realistic tissue environments.

These challenges highlight a fundamental gap: the lack of a robust, depth-resolved feedback mechanism that can operate within highly scattering media while maintaining sensitivity to fine optical features. Here, photoacoustic signals provide localization and remain stable under scattering, but lack subcellular precision. Fluorescence signals, in contrast, offer molecular sensitivity and high optical resolution but degrade rapidly with depth. Approaches that integrate both modalities could therefore leverage their complementary strengths for deep imaging through wavefront shaping and computational imaging. to garner interest in the community.

To address this, we here introduce a dual-modal photoacoustic-fluorescence (PA-FL) imaging framework that integrates both modalities within a nonlinear feedback scheme for combined wavefront shaping and computational imaging through scattering media. In this approach, the PA signal provides depth-specific guidance and coarse localization, while the fluorescence signal enables high-resolution image formation through computational reconstruction and enables high-resolution image formation within the photoacoustically defined region. Two proof-of-concept experiments validate this concept. The first demonstrates PA-FL co-guided wavefront shaping using a nonlinear hybrid feedback metric, which improves focusing stability and precision under scattering. The second establishes PA-enhanced computational fluorescence imaging, where iterative PA-guided focusing increases fluorescence generation and enables high-resolution image reconstruction within the targeted region. Together, these experiments show that integrating photoacoustic and fluorescence feedback can unify coarse localization through scattering and optical-resolution imaging within a single framework.

By uniting photoacoustic and fluorescence modalities in a single hybrid imaging system, this work establishes a potential general strategy for multimodal optical imaging that overcomes the fundamental depth-resolution limitation of optical methods. The approach combines the large depth penetration of photoacoustics with the high spatial resolution and specificity of fluorescence, offering a pathway toward deep, high-resolution imaging of complex biological tissues.

## Results

*1. Concepts of dual-modal wavefront shaping and imaging*

The working principle of the proposed approach is summarized in Figure 1. Conventional fluorescence microscopy (FLM) provides molecularly specific contrast but is limited to shallow depths, as scattering distorts both excitation and emission pathways (Fig. 1a, left). Photoacoustic microscopy (PAM), in contrast, detects ultrasound waves generated by absorbed optical energy and therefore maintains signal integrity at much greater depths, though with acoustically limited resolution (Fig. 1a, right).

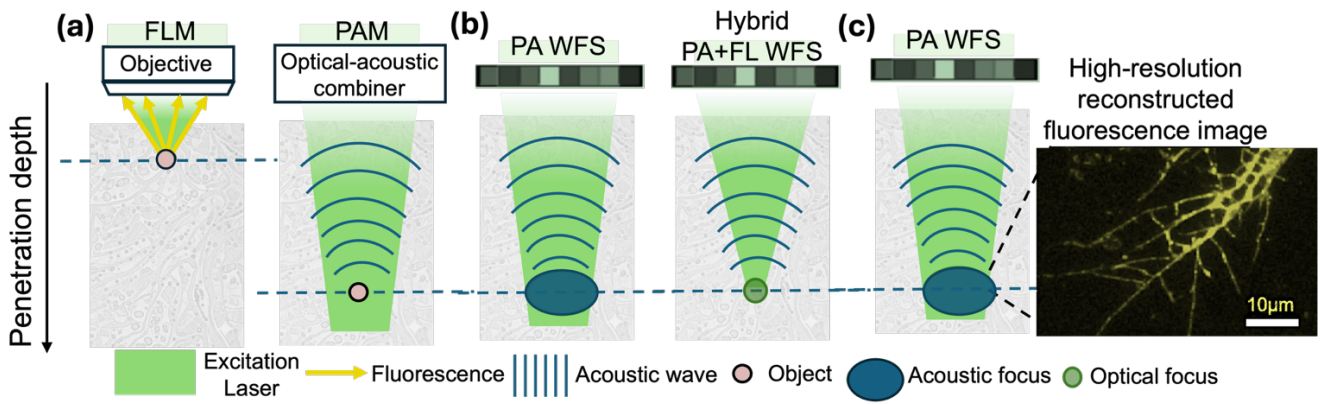

**Figure 1. Concept of hybrid photoacoustic-fluorescence wavefront shaping for deep, high-resolution imaging through scattering media**. (a) Comparison of conventional fluorescence microscopy (FLM) and photoacoustic microscopy (PAM) illustrating the depth-resolution trade-off: FLM provides subcellular resolution but is limited to shallow penetration, whereas PAM achieves deeper imaging but with coarser resolution determined by acoustic diffraction. (b) Left: schematic of photoacoustic (PA)-guided wavefront shaping (WFS), which concentrates optical energy to a depth defined by the acoustic focus but remains limited in spatial precision. Right: proposed hybrid PA + FL wavefront shaping, in which fluorescence feedback refines the optical focus within the PA-defined region, achieving both depth targeting and optical-scale resolution. (c) Proposed PA-enhanced computational fluorescence imaging. The resulting enhanced fluorescence emission, generated during PA-guided focusing, can be used for computational high-resolution image reconstruction, combining the complementary strengths of the two modalities.

Our approach aims to bridge these two regimes by combining photoacoustic and fluorescence feedback for light control through scattering media (Fig. 1b). In the first stage, photoacoustic wavefront shaping defines the approximate location of the absorber and serves as a coarse-localization guide for optical energy delivery. The photoacoustic signal provides a spatially coarse but robust feedback channel that remains stable even when the fluorescence optical signal is weak. Building on this, the hybrid optical-acoustic feedback metric integrates both signals into a nonlinear optimization scheme. While the photoacoustic response constrains the focus to the correct depth and region, the fluorescence component increases sensitivity to fine spatial variations, enabling sub-diffraction-limited refinement within the photoacoustically selected region. This dual feedback allows stable focusing even when one modality gives insufficient signal for wavefront optimization, improving both accuracy and convergence compared with single-modality control. Once optimized, the tailored wavefront concentrates excitation at the target region, enhancing fluorescence emission that can be simultaneously collected for image formation (Fig. 1c). The fluorescence generated under photoacoustic guidance contains high-resolution spatial information that can be reconstructed computationally to produce detailed images of structures embedded within scattering media and it minimizes contributions from 'outside' fluorescence which would add noise in the reconstruction. This concept unifies coarse localization at depth and fine optical resolution within a single framework. By coupling photoacoustic coarse localization with fluorescence precision, the approach establishes the foundation for two distinct, proof-of-concept demonstrations: (1) hybrid-metric wavefront shaping for robust focusing, and (2) photoacoustically guided and enhanced computational fluorescence imaging through scattering media.

*2. Multimodal system design and signal acquisition*

Building upon the concept illustrated in Fig. 1, we developed a hybrid photoacoustic-fluorescence (PA-FL) wavefront-shaping system that integrates both modalities within a single optical platform (Fig. 2a-c, for detailed setup diagram see Supplementary Figure 1). The dual-modality system is capable of acquiring co-registered photoacoustic and fluorescence signals under identical optical excitation, which serves as a basis for the nonlinear hybrid-feedback and imaging demonstrations described in the following sections.

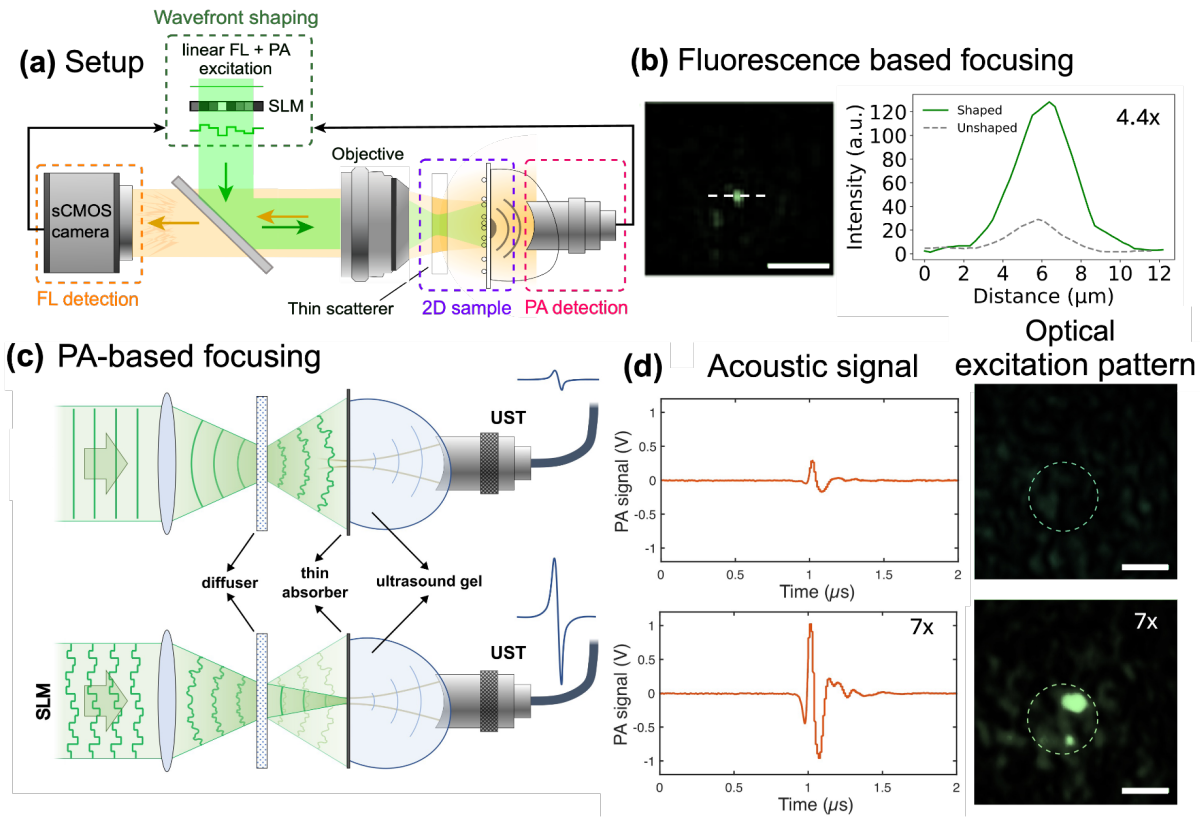

**Figure 2**. **Multimodal experimental system and single-modality wavefront shaping through scattering media.** (a) Schematic of the hybrid photoacoustic-fluorescence (PA-FL) setup. A single pulsed laser beam (532 nm) is modulated by a spatial light modulator (SLM) for wavefront shaping and directed through a thin scattering layer onto a two-dimensional sample containing absorbers and fluorophores. Fluorescence emission is collected in epi-detection by an sCMOS camera, while photoacoustic (PA) signals are recorded by an ultrasound transducer (UST) placed behind the sample. (b) Fluorescence-based focusing. Left: excitation pattern acquired through the scattering layer after fluorescence-based wavefront shaping shows convergence to a bright, diffraction-limited focus. Scale bar: 20 µm. Right: line-profile of the signal before and after shaping. (c) Photoacoustic-based focusing. Top: schematic illustrating PA feedback from a thin optical absorber through a diffuser. Bottom: comparison of acoustic waveforms and optical excitation patterns obtained without (top) and with (bottom) wavefront shaping. (d) Representative hybrid readout of PA signal and optical excitation pattern. After optimization, the PA waveform amplitude increases by approximately 7×, and the corresponding excitation intensity becomes localized within the transducer's focal region (dashed circle). Scale bar: 200 µm.

The SLM generated phase holograms to modulate the excitation wavefront. A spatial filter (aperture) in the Fourier plane of the lens selected the first diffraction order, which was re-collimated and imaged onto the back focal plane of a 10× microscope objective. The objective focused the modulated beam onto a two-dimensional sample composed of a single-layer distribution of absorbing beads and/or fluorescent beads, with different scattering media inserted between the objective and sample to emulate different levels of optical scattering.

When illuminated by the pulsed 532 nm beam, the sample generated both fluorescence emission and photoacoustic waves. Fluorescence signals were collected in an epi-detection geometry through the same objective and imaged onto a camera. Fluorescence emission was spectrally isolated from the excitation beam by a long-pass dichroic mirror and band-pass filters. Photoacoustic signals were detected by a 20 MHz focused ultrasound transducer positioned behind the sample in a water-immersion coupling medium. The acoustic signals were amplified and digitized (details in Method section).

This configuration allowed synchronous acquisition of photoacoustic and fluorescence feedback from a single excitation beam, ensuring precise spatial and temporal registration between the two signals. The setup could operate in three modes: first, fluorescence feedback mode: fluorescence intensity recorded by the camera guided iterative SLM phase optimization; second, photoacoustic feedback mode: acoustic amplitude from the transducer provided coarse-localization feedback based on optical absorption; finally, the hybrid feedback mode: both signals were simultaneously recorded and combined into a nonlinear dual-signal metric for adaptive optimization. This allows us to perform fair comparative studies between hybrid and single mode.

Example raw signals acquired through scattering media are shown in Fig. 2d and 2d. The unshaped illumination produced weak, diffuse fluorescence and broad, low-amplitude photoacoustic responses. After wavefront shaping using photoacoustic feedback, the photoacoustic amplitude increased approximately 7-fold (Fig. 2d), with the temporal waveform showing a clear single-peak signature corresponding to focused absorption at the target plane. Similarly, fluorescence optimization led to a localized excitation spot with an intensity enhancement of 4.4 relative to the background (Fig. 2b). These measurements quantitatively confirm that both modalities provide strong, independent feedback for light focusing through scattering layers.

Overall, this shows that our single-platform system enables simultaneous acquisition of co-registered photoacoustic and fluorescence signals through scattering media using a shared excitation beam. It establishes a robust experimental foundation for evaluating how nonlinear hybrid feedback enhances focusing stability (Section 3) and for demonstrating how photoacoustic-guided focusing supports high-resolution fluorescence imaging (Section 4).

*3. Nonlinear dual-signal feedback enables robust focusing*

Numerical simulations first compared the performance of fluorescence-only and hybrid feedback strategies (Figure 3a and 3b). In fluorescence-only optimization based on intensity variance, focusing was highly sensitive to initial condition and often produced arbitrary foci on any bead in the field of view, often away from the transducer location (Figure 3b). Using PA-only feedback confines the focus to an area limited by the acoustic diffraction limit. In contrast, the nonlinear hybrid metric combining fluorescence intensity variance and PA peak-to-peak signal overcame these constraints by coupling PA-based coarse localization with fluorescence-based fine adjustment (Figure 3a).

Having established the experimental platform for co-registered photoacoustic (PA) and fluorescence (FL) acquisition, we next evaluated how nonlinear hybrid feedback enhances optical focusing through scattering media. The goal was to determine whether combining both PA and FL signals into a single optimization metric could improve focusing efficiency and robustness, particularly when one modality alone does not allow focusing because of high levels of scattering or noise.

The optimization workflow used for hybrid feedback is summarized in Supplementary Figure 2. Briefly, a Hadamard-basis modulation scheme was implemented to efficiently explore the SLM phase space. Each Hadamard mode was sequentially projected onto the SLM, with discrete phase steps applied to measure the system's response. For every mode, both PA waveforms and fluorescence images were recorded simultaneously: three PA signals over a 100 ms acquisition window and fluorescence frames over the same exposure period. The PA feedback was quantified by its peak-to-peak acoustic amplitude, while fluorescence feedback was characterized by its intensity variance, which is sensitive to the formation of a bright focal spot. The two normalized signals were combined through a nonlinear weighting function that balanced PA coarse localization and fluorescence sensitivity (Hybrid metric). The phase that maximized the composite metric was applied to the cumulative SLM map, and the process repeated until convergence, typically within 70-80 iterations.

Experimental results corroborated these findings (Figure 3c). With hybrid-metric based focusing, it effectively defines a region of interest for fluorescence refinement. Within this PA-defined zone, fluorescence optimization concentrated the excitation field into a sharply confined optical focus. The temporal progression of optimization, shown in Figure 3c, reveals monotonic growth in the overall hybrid metric over successive iterations. The steady rise indicates cooperative convergence, where the PA signal stabilizes the focus position and the fluorescence response in turns refines its spatial precision. The simulations and experiments demonstrate that combining photoacoustic and fluorescence feedback substantially improves both the precision and stability of optical focusing through scattering media.

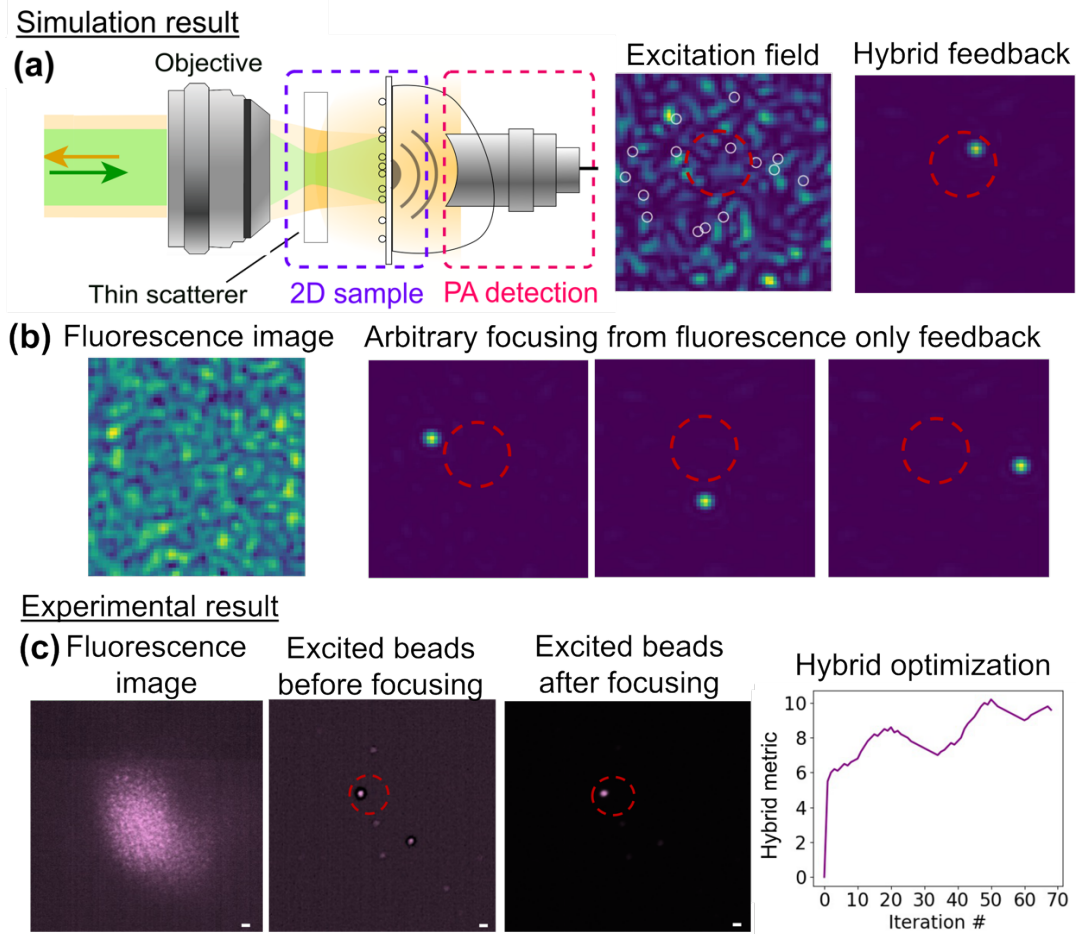

**Figure 3. Hybrid FL-PA-guided wavefront shaping enables spatially targeted focusing.** (a) Conceptual illustration of the hybrid system, where light transmitted through a thin scattering medium illuminates a 2D sample (fluorescent beads indicated by white circles in the excitation field image). Fluorescence provides a global feedback signal, while PA detection, defined by the transducer sensitivity region (red dashed circle), provides spatial guidance for targeted optimization. (b) Optimization using a fluorescence-only metric (e.g., intensity variance) leads to enhanced signal but does not guarantee localization to a predefined target region (red dashed circle), resulting in stochastic focusing at arbitrary positions. (c) Experimental results: PA-guided optimization consistently generates a localized optical focus through scattering media (5° diffuser) within the transducer detection area (red dashed circle). The optimization curve (right) shows convergence of the hybrid metric over iterations. (scale bar: 20 μm).

The hybrid metric provides localized guidance from the PA channel and optical-scale refinement from fluorescence, yielding tighter localized focusing under challenging conditions. This capability forms the basis for the next stage of experiments, where the same PA-guided focusing strategy is leveraged to generate fluorescence signals, suitable for computational high-resolution imaging through scattering media, described below.

*4. PA-guided wavefront shaping with fluorescence imaging through scattering*

Building on the robust focusing performance demonstrated with hybrid feedback, we next examined how photoacoustic-guided wavefront shaping can enhance fluorescence imaging through scattering media. The central concept is that photoacoustic feedback provides coarse localization and energy confinement, while the resulting enhanced localized fluorescence signal due to enhanced excitation, generated from the same excitation during iterative optimization process, can be harvested for computational high-resolution reconstruction within the photoacoustic-defined region.

The workflow for this PA-guided fluorescence imaging scheme is summarized in Supplementary Figure 3. The optimization again followed a Hadamard-basis phase modulation sequence, but here only the photoacoustic signal served as the feedback metric. For each Hadamard mode, the corresponding phase pattern was displayed on the SLM, and the photoacoustic response was recorded over 100 ms to compute its peak-to-peak amplitude. The phase producing the largest PA amplitude was selected and added to the cumulative SLM phase map. After convergence, the optimized phase pattern, representing the photoacoustic-defined focus, was maintained while a fluorescence image sequence was acquired. The fluorescence data were then processed through a robust computational reconstruction algorithm[39] based on principal component analysis (PCA), non-negative matrix factorization (NMF), cross-correlation and deconvolution to recover the spatial fluorescence distribution within the region of interest.

The fluorescence images acquired before and after PA-guided wavefront shaping reveal a decisive improvement in both spatial resolution and localization. In simulation, we can explore various conditions to compare image reconstruction qualities under unshaped illumination and shaped illumination. Without PA-enhanced illumination, the image quality is worse and more uncertain compared with PA-enhanced illumination (Fig. 4a). In contrast, iterative PA-guided shaping progressively localizes excitation, enabling accurate reconstruction of individual fluorescent absorbers. Experimentally, PA-guided wavefront shaping confines fluorescence to a well-localized focal spot that coincides with the acoustic detection region (Fig. 4b-d). Using fluorescent, absorbing bead targets, fluorescence reconstructions are obtained by leveraging phase-modulated fluorescence signals collected during iterative PA focusing. The results demonstrate robust localization and reconstruction across increasing scattering complexity, including a diffuser, parafilm, and ex vivo brain tissue, highlighting the ability of PA guidance to extract spatially resolved fluorescence information deep within strongly scattering media. Notably, we demonstrate high-resolution fluorescence image reconstruction through 300 μm-thick ex vivo brain slices, corresponding to approximately 5-6 optical scattering lengths at 532 nm, well beyond the ~1-2 scattering lengths over which conventional widefield fluorescence microscopy can maintain spatial fidelity.

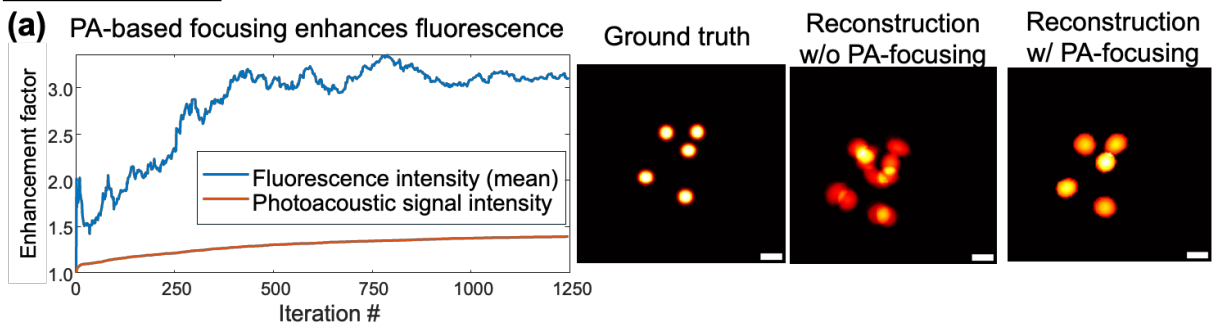
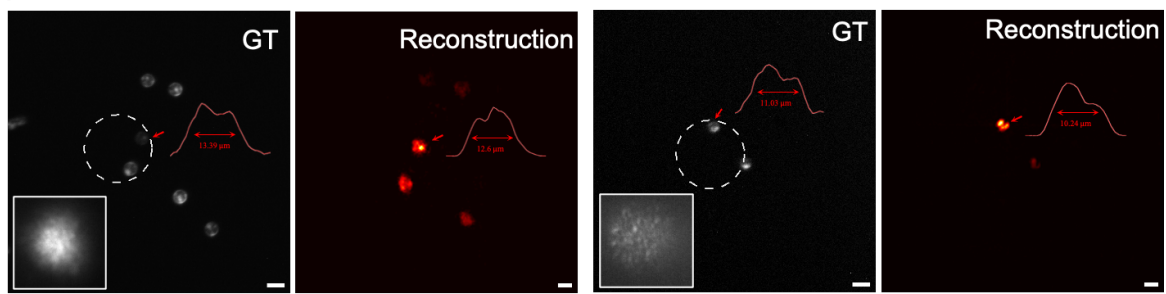
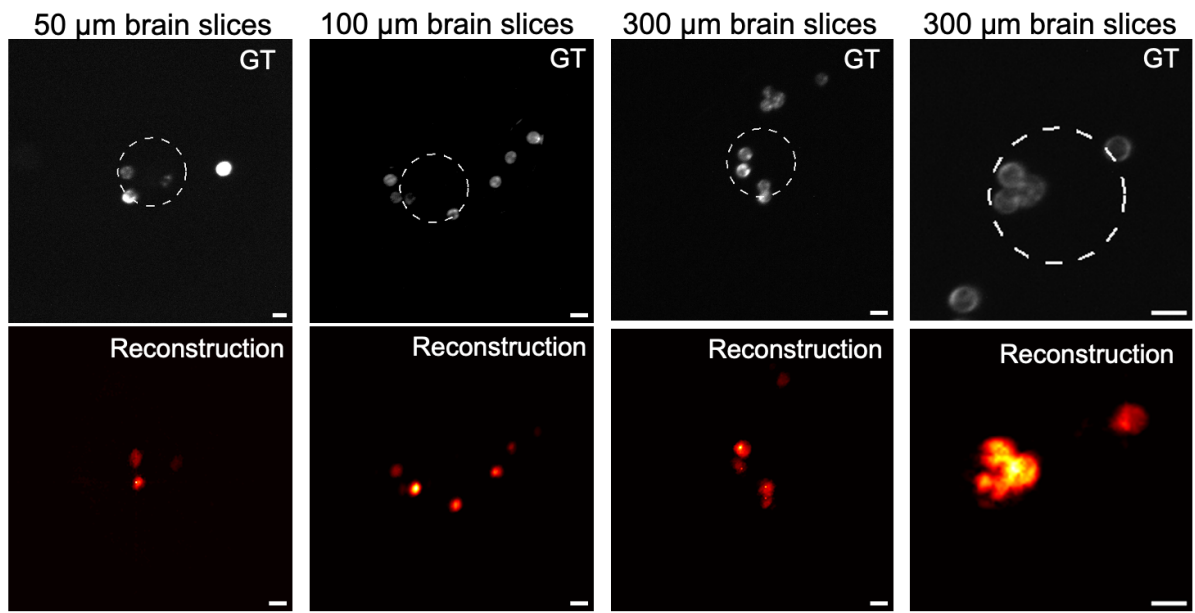

**Figure 4. Photoacoustic-guided fluorescence reconstruction through scattering media.** (a) Simulation results of reconstruction of fluorescent targets from phase-modulated fluorescence signals acquired during iterative photoacoustic (PA) focusing. PA-guided optimization progressively improves spatial confinement of excitation, enabling more accurate reconstruction compared to unguided illumination. (b-c) Experimental results of PA-enabled fluorescence (FL) imaging through scattering layers, including a 5° diffuser in (b) and a single-layer parafilm in (c). Ground-truth (GT) images and corresponding PA-guided reconstructions are shown, demonstrating successful recovery of fluorescent targets within the PA detection region (dashed circles). (d) Experimental results. PA-enabled FL imaging through scattering biological tissues (ex vivo brain slices) of increasing thickness (50 μm, 100 μm, and 300 μm). Despite strong optical scattering, PA-guided focusing enables reconstruction of embedded targets, with performance degrading gradually with tissue thickness. Fluorescent targets consist of ~10 μm absorbing beads that also emit fluorescence. All scale bars represent 20 μm.

These results confirm the benefits of the approach: the photoacoustic feedback acts as a guide that defines the coarse excitation region, while the fluorescence signal encodes fine spatial details within that region. This combination effectively merges the complementary strengths of the two modalities: the localization of acoustic wave through scattering and the high spatial precision of fluorescence. Moreover, because the fluorescence signal is collected passively during the PA-guided optimization, no additional alignment or excitation path is required, and the same optical configuration supports both focusing and imaging. Importantly, this approach extends beyond direct fluorescence enhancement—it also enables computational fluorescence imaging through scattering.

**Discussion**

Optical imaging in biological tissue remains fundamentally constrained by the depth-resolution trade-off imposed by light scattering. While photoacoustic (PA) and fluorescence (FL) imaging each address one side of this problem—depth penetration or spatial resolution—neither alone can overcome both limitations. Bridging this divide is essential for advancing high-resolution imaging in intact biological tissue, where visualizing cellular and molecular processes deep within scattering tissue remains a defining challenge in biomedicine.

Here, we have demonstrated a dual-modal photoacoustic-fluorescence (PA-FL) imaging strategy that unifies the strengths of both modalities within frameworks for dual-mode-guided wavefront shaping and augmented computational imaging. Our results establish two key advances. First, we show that PA-FL co-guided wavefront shaping enables reliable focusing under multiple scattering with a diffuser. The nonlinear dual-signal feedback metric compensates for signal degradation and enhances the controllable high-resolution localization compared to single-modality feedback approaches. Second, we demonstrate PA-enhanced fluorescence imaging, in which PA-guided wavefront shaping defines a coarse region of interest at depth while the fluorescence signal generated during the process is harvested for robust computational reconstruction beyond 5-6 scattering lengths compared to that 1-2 scattering lengths in conventional widefield fluorescence microscope. This configuration inherently couples depth targeting with optical-scale imaging, revealing a new approach for achieving high-resolution optical imaging in scenarios previously inaccessible to fluorescence methods.

Compared with prior single-modality feedback systems, our approach provides a significant leap in both stability and design flexibility. Photoacoustic-only feedback offers deep but coarse control, limited by acoustic resolution and signal sensitivity. Fluorescence-only feedback achieves fine focusing but rapidly loses signal strength in turbid media. By integrating these feedback channels nonlinearly, our dual-modal scheme leverages their complementary information content: PA defining coarse localization and FL refining focus resulting in improved focusing fidelity, robustness to noise, and enhanced imaging contrast. Beyond its immediate technical impact, this concept establishes a general framework for multimodal feedback integration, offering a foundation for future systems that combine optical, acoustic, or nonlinear contrast mechanisms for imaging through complex media.

Despite these advances, several limitations remain. The current demonstrations were conducted in controlled lab environments either in scattering phantom or in ex vivo tissue rather than in vivo, and translation to live biological systems will require improvements in SNR, temporal stability, and likely a move to more infrared wavelengths. The current iterative nature of wavefront shaping constrains imaging speed (a few seconds per iteration), which remains insufficient for capturing dynamic biological processes or living tissue motion. Additionally, deeper or more heterogeneous tissues may exacerbate scattering and acoustic attenuation, necessitating more sensitive detectors and adaptive algorithms. Managing optical power deposition while maintaining sufficient feedback for focusing will also be critical for translating this to more complex biological applications. Finally, the present implementation is limited to two-dimensional focusing; extending this approach to volumetric and time-resolved imaging represents an important step toward practical biomedical applications.

Looking forward, several avenues could substantially enhance the impact of this dual-modal paradigm. Accelerating wavefront optimization through hardware advances (e.g., high-speed spatial light modulators[44], acousto-optic deflectors[45]) and algorithmic innovation (e.g., deep learning-based phase prediction[1,2]) may enable real-time operation. Combining the method with photoswitchable fluorophores[46], nonlinear optical probes, or genetically encoded sensors could improve molecular specificity and functional imaging capabilities. Furthermore, expansion to three-dimensional, multimodal imaging, potentially integrating other contrast mechanisms such as Raman scattering or optical coherence, could provide comprehensive characterization of tissue structure and function at depth taking into account other chemical or physical information. Ultimately, coupling this dual-modal framework with in vivo models and clinical imaging systems could open new frontiers for visualizing cellular dynamics and disease progression in intact tissues.

In conclusion, this study establishes a new class of multimodal optical imaging that bridges the depth-resolution barrier by uniting photoacoustic and fluorescence modalities within a single, nonlinear feedback architecture. By exploiting their complementary physical contrasts, the approach achieves depth targeting with optical precision and robust performance under scattering conditions. As hardware and computational methods advance, this framework may transform deep-tissue optical imaging, enabling detailed, high-resolution visualization of complex biological structures and processes across scales.

## Methods

**Experimental setup.** A nanosecond pulsed laser at 532 nm (CryLas eMOPA-532; 0.5 mJ pulse energy, 400 Hz repetition rate) served as the common excitation source for both modalities. Its high pulse energy and long coherence length (~300 mm) were suitable for both unfocused photoacoustic generation and holographic phase modulation. The same laser source also allows fluorescence excitation. The output of the laser was expanded by telescopes and power-adjusted using half-wave plates and polarizing beam splitters. The beams were combined by a dichroic mirror and expanded through a telescope to overfill the display of a liquid-crystal-on-silicon spatial light modulator (SLM; Meadowlark HSP1920-500-1200-HSP8, see SI Fig. 1).

**Numerical simulations.** Numerical simulations were performed to evaluate the proposed photoacoustic-guided wavefront shaping approach under controlled scattering conditions. A coherent plane wave was modulated by a phase-only spatial light modulator and propagated through a simulated scattering medium to generate a speckle illumination pattern at the target plane. The speckle field simultaneously excited photoacoustic signals, used as feedback for wavefront optimization, and fluorescence emission, which was recorded by a virtual camera. Photoacoustic detection was modeled using a spatially localized acoustic response corresponding to the focal region of an ultrasound transducer. Wavefront optimization was carried out by sequential phase modulation of orthogonal input modes, and the optimal phase maximizing the photoacoustic signal was selected for each mode. Fluorescence image formation was simulated assuming emission proportional to the excitation intensity, with scattering-induced blurring modeled using a convolution-based approach. Further implementation flowchart is provided in the SI Figure 4.

**Computational Fluorescence Image Reconstruction Algorithm.** Image sequences acquired under iterative speckle illumination during SLM optimization are reconstructed based on a slightly custom robust non-negative principal matrix factorization (RNP)[39]. Each raw frame is first filtered and decomposed by principal component analysis into a low-rank background component and a sparse feature component, effectively suppressing static background fluorescence and enhancing temporally fluctuating speckle features. The extracted sparse feature images from all frames are assembled into a data matrix and processed using non-negative matrix factorization (NMF) to separate spatial emitter-associated components from their temporal modulation. Following factorization, pairwise deconvolution is applied to the recovered emitter components to estimate their relative spatial separations. The final image is reconstructed by assembling all emitters based on these relative distance estimates. This framework provides robust and stable reconstruction for low-contrast samples and is well matched to the intrinsic intensity fluctuations produced by diffuser rotation or SLM optimization in deep scattering environments.

**Wavefront Shaping Algorithm.** Wavefront shaping was implemented using a square active area of the spatial light modulator (SLM), which was partitioned into uniformly sized macropixels. Each macropixel consisted of either 16 × 16 or 8 × 8 native SLM pixels, defining the spatial resolution of wavefront control. An orthogonal Hadamard basis was employed to modulate the phase of the macropixels efficiently. Each Hadamard mode was sequentially addressed in the Fourier domain and modulated using a phase-stepping procedure with 4 to 8 discrete phase steps, spanning a full 0 to $2\pi$ phase cycle. The corresponding feedback signals were recorded at each phase step. Two optimization metrics were evaluated. In the photoacoustic (PA) metric, the feedback was defined as the peak-to-valley amplitude of the PA signal over the phase-stepping cycle. In the hybrid metric, the feedback was defined as the product of the PA peak-to-valley amplitude and the variance of the corresponding fluorescence signal, thereby jointly accounting for acoustic focusing efficiency and fluorescence modulation. For each Hadamard mode, the phase value that maximized the selected feedback metric was retained. This procedure was repeated for all modes to construct the optimized wavefront.

# Acknowledgements

We acknowledge the Chan Zuckerberg Initiative (Deep Tissue Imaging Grant Nos. 2020-225346 and 2024-337799)

**References**

1. Gigan, S. *et al.* Roadmap on wavefront shaping and deep imaging in complex media. *J. Phys. Photonics* **4**, 042501 (2022).

2. Xia, F. *et al.* Neurophotonics beyond the surface: unmasking the brain's complexity exploiting optical scattering. *Neurophotonics* **11**, S11510–S11510 (2024).

3. Horstmeyer, R., Ruan, H. & Yang, C. Guidestar-assisted wavefront-shaping methods for focusing light into biological tissue. *Nat. Photonics* **9**, 563–571 (2015).

4. Gigan, S. Optical microscopy aims deep. *Nat. Photonics* **11**, 14–16 (2017).

5. Wang, X. *et al.* Noninvasive laser-induced photoacoustic tomography for structural and functional in vivo imaging of the brain. *Nat. Biotechnol.* **21**, 803–806 (2003).

6. Zhang, H. F., Maslov, K., Stoica, G. & Wang, L. V. Functional photoacoustic microscopy for high-resolution and noninvasive in vivo imaging. *Nat. Biotechnol.* **24**, 848–851 (2006).

7. Wang, L. V. & Hu, S. Photoacoustic Tomography: In Vivo Imaging from Organelles to Organs. *Science* **335**, 1458–1462 (2012).

8. Yao, J. & Wang, L. V. Photoacoustic microscopy. *Laser Photonics Rev.* **7**, 758–778 (2013).

9. Wang, L. V. & Yao, J. A practical guide to photoacoustic tomography in the life sciences. *Nat. Methods* **13**, 627–638 (2016).

10. Denk, W., Strickler, J. H. & Webb, W. W. Two-Photon Laser Scanning Fluorescence Microscopy. *Science* **248**, 73–76 (1990).

11. Nie, S., Chiu, D. T. & Zare, R. N. Probing Individual Molecules with Confocal Fluorescence Microscopy. *Science* **266**, 1018–1021 (1994).

12. Lichtman, J. W. & Conchello, J.-A. Fluorescence microscopy. *Nat. Methods* **2**, 910–919 (2005).

13. Horton, N. G. *et al.* In vivo three-photon microscopy of subcortical structures within an intact mouse brain. *Nat. Photonics* **7**, 205–209 (2013).

14. Ntziachristos, V. & Razansky, D. Molecular Imaging by Means of Multispectral Optoacoustic Tomography (MSOT). *Chem. Rev.* **110**, 2783–2794 (2010).

15. Taruttis, A. & Ntziachristos, V. Advances in real-time multispectral optoacoustic imaging and its applications. *Nat. Photonics* **9**, 219–227 (2015).

16. Booth, M. J. & Patton, B. R. *Adaptive Optics for Fluorescence Microscopy*. *Fluorescence Microscopy: Super-Resolution and other Novel Techniques* 33 (2014). doi:10.1016/B978-0-12-409513-7.00002-6.

17. Antonello, J. *Optimisation-Based Wavefront Sensorless Adaptive Optics for Microscopy*.

18. Booth, M. J. Adaptive optics in microscopy. *Philos. Trans. R. Soc. Math. Phys. Eng. Sci.* **365**, 2829–2843 (2007).

19. Gould, T. J., Burke, D., Bewersdorf, J. & Booth, M. J. Adaptive optics enables 3D STED microscopy in aberrating specimens. *Opt. Express* **20**, 20998–21009 (2012).

20. Denk, W., Strickler, J. H. & Webb, W. W. Two-photon laser scanning fluorescence microscopy. *Science* **248**, 73–76 (1990).

21. Kobat, D. *et al.* Deep tissue multiphoton microscopy using longer wavelength excitation. *Opt. Express* **17**, 13354–13354 (2009).

22. Horton, N. G. *et al.* In vivo three-photon microscopy of subcortical structures within an intact mouse brain. *Nat. Photonics* **7**, 205–205 (2013).

23. Thendiyammal, A., Osnabrugge, G., Knop, T. & Vellekoop, I. M. Model-based wavefront shaping microscopy. *Opt. Lett.* **45**, 5101 (2020).

24. Popoff, S. M. *et al.* Measuring the Transmission Matrix in Optics: An Approach to the Study and Control of Light Propagation in Disordered Media. *Phys. Rev. Lett.* **104**, (2010).

25. Andreoli, D. *et al.* Deterministic control of broadband light through a multiply scattering medium via the multispectral transmission matrix. *Sci. Rep.* **5**, (2015).

26. Mounaix, M., Defienne, H. & Gigan, S. Deterministic light focusing in space and time through multiple scattering media with a time-resolved transmission matrix approach. *Phys. Rev. A* **94**, (2016).

27. Boniface, A., Dong, J. & Gigan, S. Non-invasive focusing and imaging in scattering media with a fluorescence-based transmission matrix. *Nat. Commun.* **11**, (2020).

28. Chaigne, T., Gateau, J., Katz, O., Gigan, S. & Bossy, E. Photoacoustics-assisted wavefront shaping in turbid media, and improved photoacoustics imaging exploiting light coherence properties. in *Optics InfoBase Conference Papers* 1–1 (2016). doi:10.1364/OTS.2016.OTu2A.2.

29. Inzunza-Ibarra, M. A., Premillieu, E., Grünsteidl, C., Piestun, R. & Murray, T. W. Sub-acoustic resolution optical focusing through scattering using photoacoustic fluctuation guided wavefront shaping. *Opt. Express* **28**, 9823 (2020).

30. Sun, J. *et al.* Photoacoustic Wavefront Shaping with High Signal to Noise Ratio for Light Focusing Through Scattering Media. *Sci. Rep.* **9**, 4328 (2019).

31. Xia, F., Leite, I., Prevedel, R. & Chaigne, T. Optical wavefront shaping in deep tissue using photoacoustic feedback. *J. Phys. Photonics* **6**, 043005 (2024).

32. Chaigne, T., Gateau, J., Katz, O., Bossy, E. & Gigan, S. Light focusing and two-dimensional imaging through scattering media using the photoacoustic transmission matrix with an ultrasound array. *Opt. Lett.* **39**, 2664 (2014).

33. Vellekoop, I. M. & Mosk, A. P. Focusing coherent light through opaque strongly scattering media. *Opt. Lett.* **32**, 2309–2311 (2007).

34. Akbulut, D., Huisman, T. J., Putten, E. G. van, Vos, W. L. & Mosk, A. P. Focusing light through random photonic media by binary amplitude modulation. *Opt. Express* **19**, 4017–4029 (2011).

35. Blochet, B., Bourdieu, L. & Gigan, S. Focusing light through dynamical samples using fast continuous wavefront optimization. *Opt. Lett.* **42**, 4994 (2017).

36. Boniface, A., Blochet, B., Dong, J. & Gigan, S. Noninvasive light focusing in scattering media using speckle variance optimization. *Optica* **6**, 1381–1385 (2019).

37. Moretti, C. & Gigan, S. Readout of fluorescence functional signals through highly scattering tissue. *Nat. Photonics* **14**, 361–364 (2020).

38. Zhu, L. *et al.* Large field-of-view non-invasive imaging through scattering layers using fluctuating random illumination. *Nat. Commun.* **13**, 1447 (2022).

39. Gao, Z. *et al.* Fluorescence microscopy through scattering media with robust matrix factorization. *Cell Rep. Methods* **5**, 101031 (2025).

40. Katz, O., Heidmann, P., Fink, M. & Gigan, S. Non-invasive single-shot imaging through scattering layers and around corners via speckle correlations. *Nat. Photonics* **8**, 784–790 (2014).

41. Wu, T., Baek, Y., Xia, F., Gigan, S. & De Aguiar, H. B. Replica-Assisted Super-Resolution Fluorescence Imaging in Scattering Media. *ACS Photonics* **12**, 1308–1315 (2025).

42. d'Arco, A., Xia, F., Boniface, A., Dong, J. & Gigan, S. Physics-based neural network for non-invasive control of coherent light in scattering media. *Opt. Express* **30**, 30845 (2022).

43. Beaurepaire, E. & Mertz, J. Epifluorescence collection in two-photon microscopy. *Appl. Opt.* **41**, 5376–5382 (2002).

44. Rocha, J. C. A. *et al.* Fast and light-efficient wavefront shaping with a MEMS phase-only light modulator. *Opt. Express* **32**, 43300 (2024).

45. Blochet, B., Akemann, W., Gigan, S. & Bourdieu, L. Fast wavefront shaping for two-photon brain imaging with multipatch correction. *Proc. Natl. Acad. Sci.* **120**, e2305593120 (2023).

46. Deo, C. Bridging Light and Sound: a Spironaphtopyran-Rhodamine Dyad with High-Contrast Photoswitching Between Fluorescence and Photoacoustic Signal.

# Supplementary Information

Supplementary Figure 1

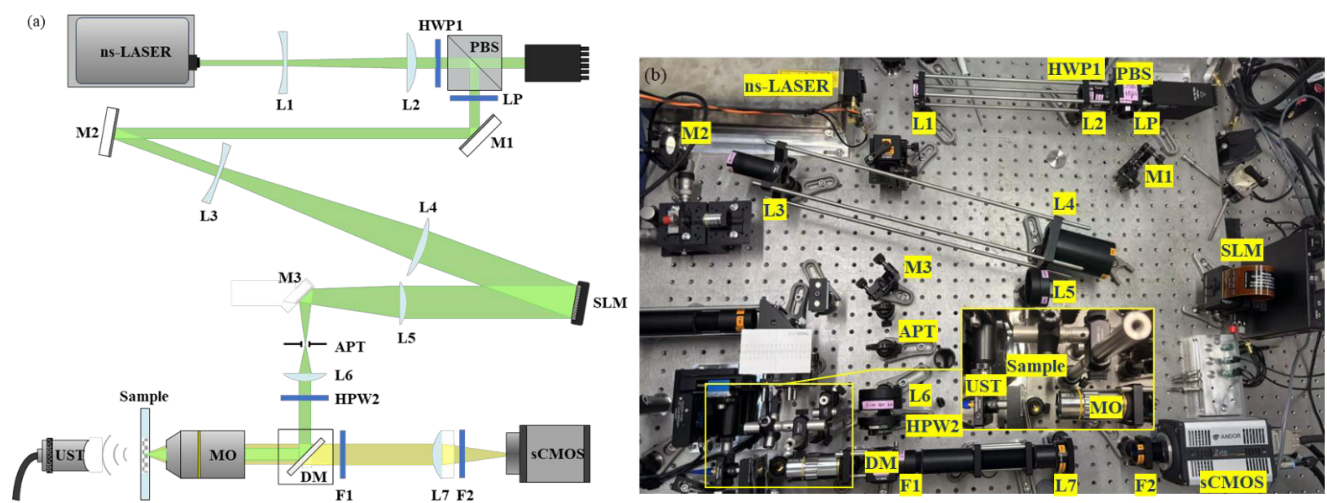

***Supplementary Figure 1.*** **Hybrid imaging system combining photoacoustics and fluorescence modalities. a,** Schematic of the complete optical system. **b,c,** Photographs of the setup assembled. The outputs from laser sources ns-LASER is expanded through two telescopes (formed by lenses L1-L2 for one beam, L3-L4 for the other, and L5-L6 for both) to overfill spatial light modulator SLM. The carrier frequency applied to the holograms displayed by the SLM aligns the 1st diffraction order of one of the beams with aperture APT in the Fourier plane of lens L7. A sample placed at the focal plane of MO emits both fluorescence and acoustic signals when excited by ns-pulsed light. Ultrasound transducer UST detects photoacoustic signals propagating from the sample through an acoustic immersion medium (not shown). MO together with tube lens L9 image the sample plane onto an sCMOS camera. Dichroic mirror DM2 and filters F1-F3 isolate the fluorescence emission signal. A diffuser is inserted between MO and the sample (not shown). *Legend:* **ns-LASER**, ns-pulsed laser source emitting at 532 nm (CryLas eMOPA-532); **SLM**, spatial light modulator (Meadowlark Optics HSP1920-500-1200-HSP8); **MO**, microscope objective (10X Mitutoyo Plan Apo Infinity Corrected Long WD Objective); **UST**, ultrasound transducer (Olympus V317-SM, 12.7 mm focal length); **sCMOS**, scientific complementary metal-oxide semiconductor camera (Andor Zyla 4.2 Plus); **M1-M2**, broadband dielectric mirrors (Newport 10D620BD.1); **M3**, D-shaped mirror; PBS, polarizing beamsplitter cubes (Thorlabs PBS251); **HWP**, half-wave plates (HWP1: Newport 10RP02.16, HWP2: WPH10ME-532); **LP**, linear polarizer; **DM**, long-pass dichroic mirror (Semrock Di03-R561-t3); **APT**, aperture (iris diaphragm), **F1**, long-pass dichroic filter (Edgebasic LP 561); **F2**, long-pass filter (RazorEdge LP 561); **L1**, plano-concave lens (f = -50 mm, Newport SPC034AR.14); **L2**, plano-convex lens (f = 300 mm, Newport KPX112.AR14); **L3**, plano-concave lens (f = -200 mm, Newport KPC028AR.14); **L4**, plano-convex lens (f = 750 mm, Newport KPX214AR.14); **L5**, achromatic doublet lens (f = 250 mm, Edmund Optics 49287); **L6**, achromatic doublet lens (f = 120 mm, S&H); **L7**, achromatic doublet lens (f = 165 mm, Zeiss 425308).

Supplementary Figure 2

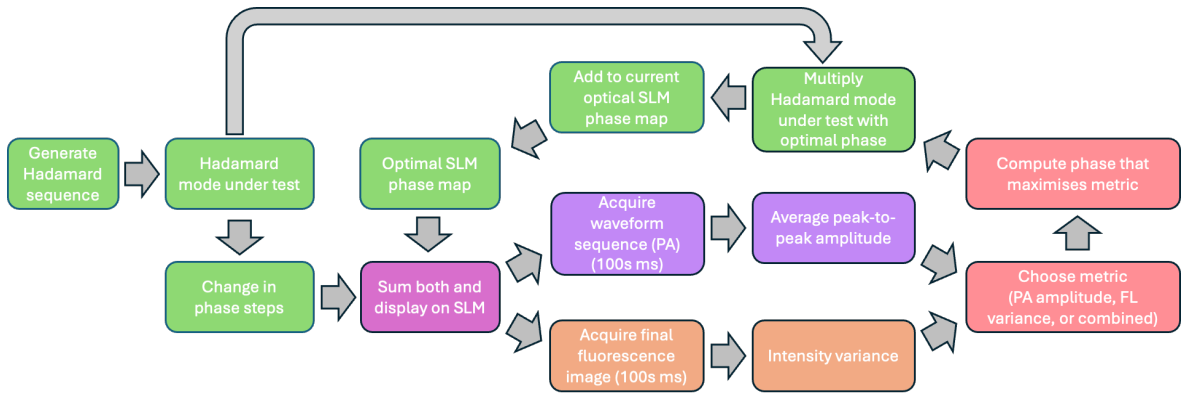

*Supplementary Figure 2*. Flowchart for hybrid wavefront shaping combining photoacoustics and fluorescence modalities

Supplementary Figure 3

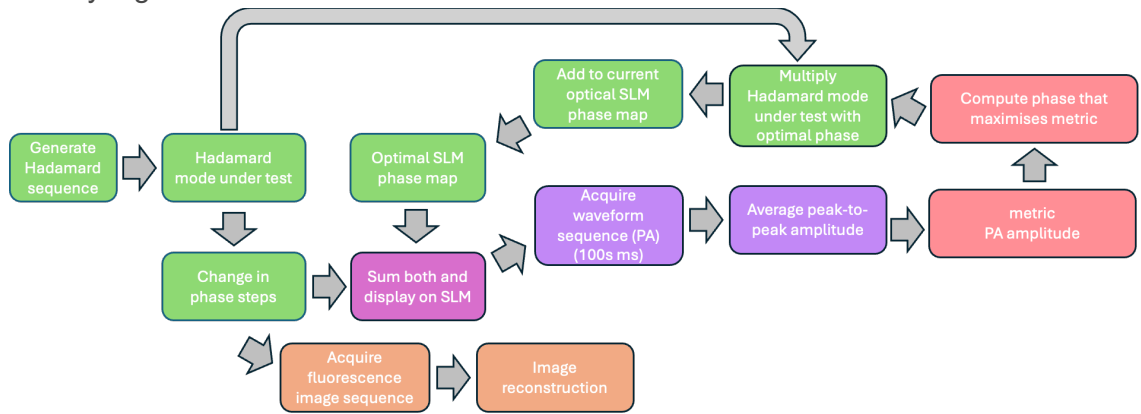

*Supplementary Figure 3*. Flowchart for PA-enhanced fluorescence imaging through scattering

Supplementary figure 4

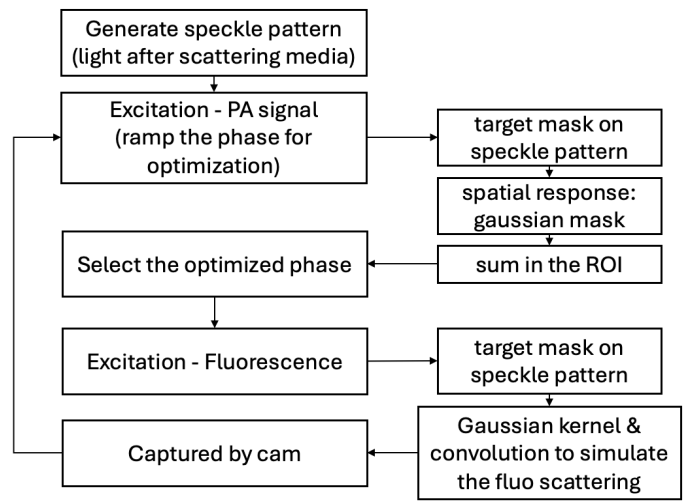

*Supplementary Figure 4*. Flowchart for simulation